\documentstyle[aps,prb,preprint]{revtex}
\begin{document}
\draft
\title
{Avalanches and the Renormalization Group for Pinned Charge-Density Waves}
\author{Onuttom Narayan}
\address
{Department of Physics, Harvard University, Cambridge, MA 02138\\{\rm and}\\
AT\&T Bell Laboratories, 600 Mountain Avenue, Murray Hill, NJ 07974}
\author{A. Alan Middleton}
\address
{NEC Research Institute, 4 Independence Way, Princeton, NJ 08540
}
\date{May 12th, 1993}
\maketitle

\begin{abstract}
The critical behavior of charge-density waves (CDWs) in the pinned phase
is studied for applied fields increasing toward the threshold field, using
recently developed renormalization group techniques and simulations of
automaton models. Despite the existence of many metastable states in the
pinned state of the CDW, the renormalization group treatment can be used
successfully to find the divergences in the polarization and the
correlation length, and, to first order in an $\epsilon = 4-d$ expansion,
the diverging time scale. The automaton models studied are a charge-density
wave model and a ``sandpile'' model with periodic boundary conditions;
these models are found to have the same critical behavior,
associated with diverging avalanche sizes. The numerical
results for the polarization and the diverging length and time scales
in dimensions $d=2,3$ are in agreement with the analytical treatment. These
results clarify the connections between the behaviour above and below
threshold: the characteristic correlation lengths on both sides of the
transition diverge with different exponents.
The scaling of the distribution of avalanches on the approach to threshold
is found to be different for automaton
and continuous-variable models.

\end{abstract}

\section{Introduction}

The static and dynamic behaviour of sliding charge density waves (CDWs)
is perhaps the most well studied example of a class of problems involving
the transport of an elastic medium through a  disordered background.
The CDW, which behaves like an elastic medium, is pinned by impurities
distributed randomly throughout the material. As the magnitude of an
externally applied electric field is varied, a depinning transition is seen,
{}from a stationary phase at weak fields, to a moving phase at strong fields
where
the CDW slides through the material. In the vicinity of this depinning
transition, the dynamics of the CDW are correlated over long distances, with
characteristic correlation lengths diverging at the threshold field. It has
been shown\cite{Fisher} that the behaviour near the threshold field can
be studied as a critical phenomenon associated with a second order phase
transition.\cite{foot}

Extensive numerical simulations\cite{Oldnum,Sibani,M1a,M1b,Myers,Automaton} on
a
number of classical models for CDWs
have helped in understanding the critical properties in the vicinity of
the depinning transition. In particular, critical exponents describing the
scaling of physical quantities can be determined for
various spatial dimensions of the CDW. Recently, it has also proved possible
to obtain
analytically the behaviour in the moving phase above threshold, through a
renormalization group treatment.\cite{N1}  The results, obtained within an
$\epsilon$-expansion for $d=4-\epsilon$ spatial dimensions, agree very well
with the numerical simulations.

While the critical behaviour above threshold is  fairly well understood,
it is more difficult to carry out an analytical treatment below threshold.
The reason why the analysis is simpler in the moving phase is that, for
models in which the elastic interactions within the CDW are strictly
convex, and no dislocations (``phase slips'') are allowed, the system
approaches a unique steady-state configuration at long times (up to time
translations).\cite{M2} A perturbation expansion around some relatively
simple configuration is
thus more likely to succeed. In the stationary phase, on the other
hand, the impurities can pin the CDW in any of several different states;
the particular state that the CDW reaches depends on the past history of
the system. Any analytical
treatment must include this history dependence, instead of being the simple
search for a single attractor, as is sufficient above threshold.

The history dependence of the CDW below threshold is most easily seen in
the polarizability. If the driving force $F$ (produced by the external
electric field) is
raised monotonically towards the depinning threshold $F_T$, the
response $\chi^\uparrow(F)$ to
infinitesimal increases in the force becomes larger and larger, and diverges at
$F_T$ as $\chi^\uparrow(F)\sim(F_T-F)^{-\gamma}$, with an exponent
$\gamma$. Also, at any  value of $F$, the response of the CDW
to increasing the force further is dominated by localized regions of activity,
or
``avalanches'', whose characteristic size can be used to define a diverging
correlation length $\xi\sim(F_T-F)^{-\nu}$.
If instead of this monotonic approach, the force is {\it lowered\/}
{}from just below threshold, the polarizability $\chi^\downarrow(F)$ does not
diverge at $F_T$ (although, depending on the specific model used for
the CDW, it  may have a cusp singularity). It is thus necessary to distinguish
between the upwards polarizability, $\chi^\uparrow(F)$, and the downwards
polarizability $\chi^\downarrow(F)$.

In this paper, we show how it is possible to obtain analytically the critical
behaviour for the `natural' monotonic approach to threshold, through
an expansion around the mean field solution with the proper history
dependence. Our results are in good agreement with recent numerical
simulations, as well as new data that we present here. We find that the
correlation length exponent is given by
\begin{equation}
\nu=2/d\label{intronu}
\end{equation}
for a $d$ dimensional system, while the polarizability exponent is
\begin{equation}
\gamma=4/d.\label{introgamma}
\end{equation}
These exponents can in fact be obtained for a highly simplified CDW
model\cite{Parisi} that can be solved exactly. However, the dynamic
exponent $z$ {\it cannot\/} be obtained {}from this model: we show that
$z$ is equal to its value above threshold
\begin{equation}
z=2-\epsilon/3+O(\epsilon^2)\label{introz}
\end{equation}
in $d=4-\epsilon$ dimensions. Later in this paper, we discuss how the
model of Ref.\onlinecite{Parisi} corresponds to a pathological limit, and
why, despite this, it obtains $\nu$ and $\gamma$ correctly.

The solution
that we obtain is similar in form to the solution above threshold.\cite{N1}
However, there are complications with two-sided scaling. This is because,
as was  seen above threshold, there is an extra relevant operator at
the fixed point of the renormalization group, which affects only the
statics and not the dynamics. This operator has very different consequences
in the pinned and moving phases of the CDW, resulting in different correlation
length exponents on the two sides of the phase transition. The absence
of two-sided scaling  is unlike the behaviour for phase transitions
in equilibrium systems, and conventional arguments about the properties of
scaling functions must be used with caution here.

In ``phase-only'' models for CDWs, the CDW is described by a periodic
modulation in the density of electrons in the material; the phase of
this modulation is assumed to be the only dynamical variable of importance
and is assumed to be a continuous function of position. In numerical
simulations, the system is discretized to a lattice, with a continuous
phase variable at each site. Although simulations on such continuous
variable models have been performed\cite{Sibani,M1a,M1b}, it is much
more efficient to simplify the model, with the phase of the CDW at any
lattice site  restricted to being an integer multiple of $2\pi$ away
{}from a (site dependent) preferred phase, and time being incremented in
discrete steps.\cite{Myers,Automaton} The resulting automaton models,
which differ in details of how these approximations are made, yield far
more precise results. Within the numerical uncertainties, much of the
critical behaviour is the same for both classes of models.
In particular, the correlation length exponent $\nu$ and
(with larger numerical uncertainties) the polarizability exponent
$\gamma$ appear to be the same. An important exception to this is the
downward polarizability,  $\chi^\downarrow$: as already mentioned,
the nature (or even the existence) of a cusp singularity in
$\chi^\downarrow$ {\it does\/} depend on the
details of the model used, and is in fact even different in different
continuous variable models. However, this difference is not seen in
the monotonic approach, which is what we shall be interested in here.

Since the approach to threshold causes the system to become increasingly
unstable, with larger and larger avalanches and a divergent polarizability,
it is perhaps not surprising that differences between automaton and
continuous dynamics do not affect the critical behaviour. An important
exception to this rule is the distribution for the {\it number\/} of
avalanches. For continuous dynamics, an infinitesimal increase in $F$ triggers
avalanches in which the change in the phase at any point is bounded above
by $2\pi$ (corresponding to unity in the automaton models). Most of the
avalanche area advances by almost $2\pi$. In large avalanches, it is highly
probable that a subsequent small increase in $F$ will ``retrigger'' avalanches
in the same region, as the original unstable point is likely to be nearly
unstable at the completion of the first avalanche. Avalanches of usually
decreasing size will be retriggered until the region is more stable.
In the automaton models, owing to the discretization of
the CDW phases, several ``retriggered'' avalanches are grouped together into a
single large avalanche. This difference between the two classes of models is
one that involves the behaviour at {\it short times\/}; this does not alter
long-wavelength low-frequency characteristics that determine $\nu$ and $z$
(nor $\gamma$, which involves a spatial average over the entire system).
However, it may affect avalanche distributions, which do depend on how
avalanches are grouped.
In this paper we present numerical data on the distribution of avalanches
for the automaton models, and examine the differences with results for
continuous variable models.

The automaton models for CDWs are closely related to some of the sandpile
models that have been proposed and studied recently.\cite{BTW} We find that the
avalanche
distribution that we obtain here for the CDW automaton models agrees
very well with the numerical results on sandpiles at the
critical point.\cite{alanref} The dynamic exponent,
$z$, that relates the characteristic duration of an avalanche to its
size, is also found to be the same numerically for both classes of models:
\begin{eqnarray}
z(d=2)&=&1.32\pm 0.04\nonumber\\
z(d=3)&=&1.65\pm 0.06\label{numz}
\end{eqnarray}
The {\it approach\/} to threshold for sandpile models has been treated
through numerics and scaling arguments by Tang and Bak\cite{Tang}; our
results differ in certain respects {}from theirs, which we discuss in
detail later in this paper.

The connection between sandpiles and CDW models can be exploited to
obtain the dynamic exponent $z$ for two dimensional systems.
 Majumdar and Dhar\cite{Majumdar} have shown that this exponent is equal to
$5/4$ for sandpile models in $d=2$; this is in fair agreement with
the numerical
results for CDWs.\cite{F2} They are also able to relate the two exponents that
enter the avalanche distribution to each other, thus leaving only one
of the two to be determined numerically. In this paper we show that
it is in fact possible to obtain the same exponent identity {}from
completely different scaling arguments on CDWs, if we assume that the total
rate of avalanche generation (which is dominated by {\it small\/} avalanches)
is not singular as threshold is approached. (We have verified this assumption
numerically for the automaton models.)

The rest of this paper is organized in the following manner. Section II
carries out the analytical treatment of the behaviour below threshold,
and obtains the critical exponents. Section III examines the
numerical results, which agree quite well with the analytical predictions.
Section IV compares the distribution of avalanches for CDW continuous
dynamics, CDW automata, and sandpile models, as well as exploring other
connections to sandpile models. Section V discusses the relationship of
this work to earlier results, and its possible relevance to other
physical systems.

\section{Analytical Results.}

The model that we use in this section is the Fukuyama-Lee-Rice
model,\cite{Fukuyama} which assumes that the distortions of the CDW are
continuous ({\it i.e.\/} that there are no dislocations). The dynamics can then
be expressed in terms of a phase variable, $\phi(x;t)$, which measures the
distortions with respect to an ideal undistorted CDW. (The position $x$ is a
$d$-dimensional vector.) Assuming that the dynamics are strongly overdamped,
and are given by a simple relaxation of an energy functional $H$, the equation
of motion is
\begin{equation}
\partial\phi(x;t)/\partial t = -\partial H/\partial\phi
=\nabla^2\phi(x;t)+F+h(x)Y(\phi-\beta(x)).\label{FLR}
\end{equation}
In Eq.(\ref{FLR}) there are three terms on the right hand side: a simple
elastic force that arises {}from an elastic energy ${1\over 2}(\nabla\phi)^2$,
a uniform force $F$ {}from the external electric field,
and a force $h(x)Y(\phi-\beta(x))$ {}from the impurity pinning.
This impurity force has an explicit dependence on the position $x$,
arising {}from the quenched randomness in the location of the impurities,
through the functions $h(x)$ and $\beta(x)$. These correspond physically
to the  strength of the impurity pinning, $h(x)$, and
a preferred phase $\beta(x)$ selected by the impurities. (In order to fix the
normalization of $h$, we choose $|Y|$ to have a maximum value of 1.)
Since the impurities have only short range correlations, $h(x)$ and
$\beta(x)$ are taken to be uncorrelated {}from one position
to another, with a distribution $\rho(h)$ for $h$ and a uniform distribution
over $[0,2\pi)$ for $\beta$.
The pinning force is the derivative, $-\partial_{\phi}V(\phi;x)$ of
an impurity pinning potential. Because of the $2\pi$ translational invariance
of the CDW, the force $Y$ has to be $2\pi$-periodic in $\phi$.

Among the quantities of physical interest in this model below threshold are the
polarization,
$P$, defined as the mean displacement {}from some initial configuration
of the phase $\phi$; close to threshold, the singular part of $P$ scales as
\begin{equation}
P_s(F)=[\langle\phi\rangle-\langle\phi\rangle_{\rm initial}]_s
\sim(F_T-F)^{-\gamma+1}.\label{polarz}
\end{equation}
The derivative of this with respect to $F$ is the upwards polarizability,
defined earlier as the response to an infinitesimal increase in $F$, which
scales as
\begin{equation}
\chi^\uparrow(F)\equiv\lim_{\Delta F\rightarrow 0^+}[P(F+\Delta F)-P(F)]/\Delta
F\sim (F_T-F)^{-\gamma}.\label{suscep}
\end{equation}

In Ref.\onlinecite{N1}, the behaviour above threshold for models of this type
was
analyzed. This was done by expanding around the solution of
Eq.(\ref{FLR}) within a mean field approximation, which is obtained by
replacing the short-ranged elastic term $\nabla^2\phi$
with an infinite-ranged interaction, $\langle\phi\rangle-\phi$. The
mean field solution is known to depend on the details of the pinning
potential $V(\phi)$: for instance, the velocity above threshold scales
as $v\sim (F-F_T)^\beta$ in the critical regime,\cite{convention} with an
exponent
$\beta_{MF}=3/2$ for smooth pinning potentials\cite{Fisher} and
$\beta_{MF}=1$ for
pinning potentials which have wedge-shaped linear cusped maxima.\cite{M1b} This
lack of universality arises {}from the fact that, for smooth pinning
potentials, when the phase $\phi$ at any site passes through an unstable
point in its local effective potential, it spends a long time accelerating
before jumping forward to a new stable state. This results in a second
time scale that diverges at threshold (in addition to the natural scale
of $2\pi/v$). No such second divergent time scale is present for linear
cusped potentials, for which the acceleration time remains of $O(1)$
arbitrarily close to threshold. Despite this difference in behaviour
for the two types of potentials in mean field theory, numerically it
is found that the critical behaviour is {\it independent\/}
of the shape of the pinning potential for $d<4$, and is in fact even the same
for automaton models. In Ref.\onlinecite{N1}, it was argued that this is
because the
dynamics for small $d$ are very irregular: when the phase at any site  jumps
forward, it produces a sharp force on all the neighbours to which it is
elastically coupled. For large $d$, the total elastic force acting on any phase
is the sum of contributions {}from a large number of neighbours; however, as
$d$ is decreased, the elastic force becomes more and more jerky. For
sufficiently small $d$, the phase $\phi$ of the CDW
at any site is `kicked' by its neighbours over the maximum of the local
effective potential it experiences. The details of the pinning potential
are no longer important, and the second divergent time scale is
eliminated, restoring universality. The expansion around mean field theory
was accordingly carried out for linear cusped potentials, for which the
second time scale is absent initially, so that the expansion
is better behaved.

When $F_T$ is approached monotonically {}from below threshold, the critical
behaviour is controlled by large avalanches, in which the CDW phase in
localized regions moves forward abruptly in response to a small change in the
field. Since the collapse to universality above threshold was understood in
terms of the jerky dynamics, the same should be true even below threshold.
This is indeed found to be the case numerically. Thus even below threshold, we
seek to expand around the mean field solution for linear cusped potentials.

Before proceeding with the calculations, it is necessary to consider a
possibility that might make this procedure erroneous. While we have
answered the question of history dependence by choosing a specific approach
to threshold, namely increasing $F$ monotonically, this does not completely
eliminate problems that arise
{}from the existence of many metastable states. An expansion around the mean
field solution of the type we develop here corresponds to
first increasing the force within an infinite ranged model, and then
tuning the elastic coupling to a short ranged form ({\it i.e.\/}, restoring the
$\nabla^2\phi$ term in Eq.(\ref{FLR}) which was replaced by $\langle\phi\rangle
-\phi$ in mean field theory). On the other hand, the physically relevant
critical behaviour corresponds to {\it first\/} tuning the elastic
coupling to a short ranged form, and {\it then\/} increasing the force
{\it within\/} a short ranged model. When there are many metastable
solutions, as is the case below threshold, the two methods could in
principle yield different critical behaviour.

Since the CDW phase at any site moves only forward when the force is raised
monotonically, such problems {}from
metastability could arise if reducing the range of the elastic interactions
{}from a mean field theory were to result in the phases at different sites
trying to move {\it backwards\/}. While this will of course be true for
{\it some\/} of the phases, for which the coupling to their neighbours
tends to retard their motion, we shall see later in this paper that the
polarizability exponent $\gamma$ is in fact larger for $d<4$
than in mean field theory. Thus the dominant effect of
reducing the range of the elastic interactions is to make the phases
move {\it forward\/}, making the polarization (and hence the polarizability)
more divergent than in mean field theory. We therefore expect our
procedure of expanding around the mean field solution to yield the correct
critical behaviour.

The expansion around mean field theory is carried out following the
prescription of Sompolinsky and Zippelius;\cite{sompol} the details are given
in Ref.\onlinecite{N1}.
The method yields an impurity averaged generating functional for the
correlation and response functions of the phase $\phi$, of the form
\begin{eqnarray}
\overline Z=\int[d\Phi]&[&d\hat\Phi]\exp\{-\int
d^d x dt\,\hat\Phi(x,t)[\partial_t-\nabla^2+r]\Phi(x,t)+\hat\Phi(x,t)
[F-F_{MF}(\langle\phi(t)\rangle)]\qquad\nonumber\\
 &-& \int d^dx dt_1 dt_2\,\hat\Phi(x,t_1)
\hat\Phi(x,t_2) C[\langle\phi(t_1)\rangle+\Phi(x,t_1)-\langle\phi(t_2)\rangle
-\Phi(x,t_2)]\}.\label{genfunc}
\end{eqnarray}
Here $\Phi$ and $\hat\Phi$ are dummy field variables, in terms of which the
long wavelength low frequency forms of the truncated correlation and response
functions of $\phi$ are given by
\begin{eqnarray}
\partial^n\langle\phi(x_1,t_1)\ldots\phi(x_m&,&t_m)
\rangle_{\rm trunc}/\partial\varepsilon(x_{m+1},t_{m+1})
\ldots\partial\varepsilon(x_{m+n},
t_{m+n})\qquad\qquad\qquad\qquad\nonumber\\&=&\langle\Phi(x_1,t_1)
\ldots\Phi(x_m,t_m)\hat\Phi(x_{m+1},t_{m+1})\hat\Phi(x_{m+n},t_{m+n})
\rangle\label{mapping}
\end{eqnarray}
where the left hand side
represents the generalized response to a perturbing force $\varepsilon(x,t)$
added to the right hand side of Eq.(\ref{FLR}), and the right hand side is the
expectation value using the measure of Eq.(\ref{genfunc}). $\Phi(x,t)$ is like
a coarse grained version of the phase $\phi(x,t)$. It
gives the deviations
of $\phi(x,t)$ {}from its spatially averaged value, $\langle\phi\rangle$,
and must satisfy the consistency condition $\langle\Phi\rangle=0$, which
determines $F-F_{MF}$. Various other terms,
that do not affect the critical behaviour, have been suppressed in
Eq.(\ref{genfunc}).

Eq.(\ref{genfunc}) is similar to the generating functional obtained in
Ref.\onlinecite{N1} above threshold, except for new terms that we discuss in
the
next paragraph.
The function $C$ is the same as the mean field phase phase correlations
as a function of time {\it above\/} threshold:
$C(vt_1-vt_2)=\langle[\phi(t_1)-vt_1][\phi(t_2)-vt_2]\rangle$. Below threshold,
the spatial average of the phase, $\langle\phi(t)\rangle$, replaces $vt$.
For any value of $\langle\phi\rangle$, the function
$F_{MF}(\langle\phi\rangle)$ is the force that would result in an average
displacement $\langle\phi\rangle$ in the mean field approximation to
Eq.(\ref{FLR}), while the function $F$ is the force that would produce the
same average displacement in the full short-ranged model of Eq.(\ref{FLR}).
As mentioned in the previous paragraph, the difference between these two,
which is the coefficient of the second term in Eq.(\ref{genfunc}), can be
found by the consistency condition $\langle\Phi\rangle=0$. This again is
similar to the case above threshold.

Despite the similarities, there are important differences between
Eq.(\ref{genfunc}) and the corresponding generating functional above
threshold\cite{N1}. There is a
non-zero `mass term' in Eq.(\ref{genfunc}), $r\hat\Phi\Phi$, which results in
a finite response of $\phi$ to a spatially uniform dc force. Above threshold,
where this term vanishes exactly due to time translational invariance
of the system, $\chi(q=0,\omega)$ has a $1/i\omega$ singularity,
corresponding to a finite response in the {\it velocity\/} of $\phi$.
The bare value of $r$ is the inverse of the mean field polarizability
$\chi^\uparrow(F)$ (along the monotonically increasing path); $r$ vanishes at
threshold, although the manner in which it approaches zero at $F_T$
depends on the distribution $\rho(h)$ of the pinning strengths. The generating
functional $\overline Z$ also differs {}from the form above threshold by
various other terms that also vanish at threshold like $\hat\Phi r\Phi$
does.  For instance, the coefficient of the second term,
$F-F_{MF}(\langle\phi\rangle)$ has corrections of $O([F_T-F]^2)$. These,
however, do not affect the critical behaviour and are not
shown in Eq.(\ref{genfunc}). Also, it is only when the mean field correlation
and response functions are evaluated within the monotonically
increasing approach to threshold that we obtain the functional forms in
Eq.(\ref{genfunc}) that resemble those above threshold. For other
(non-monotonic) approaches, this will not be the case. Above threshold,
where the steady state behaviour is unique, no particular path needs to be
specified.

The variable $\langle\phi(t)\rangle$ in the arguments of the functions
$F_{MF}$ and $C$ in Eq.(\ref{genfunc}) implicitly controls the dependence on
$F$. It is necessary to increase $\langle\phi(t)\rangle$ {\it adiabatically\/}
in order for the form in Eq.(\ref{genfunc}) to be valid; otherwise, there would
be various transients that would have to be included in the equation. Since
the increase of the force is much slower than all the dynamics of
the CDW, the equal time response and correlation functions really give the
{\it static\/} behaviour. Although it would seem that Eqs.(\ref{genfunc}) and
(\ref{mapping}) can be used to obtain responses to time dependent
perturbations, there is an implicit constraint that the perturbing force
must satisfy for our entire approach to be valid: $\varepsilon(t_1)\geq
\varepsilon(t_2)$ for $t_1\geq t_2$. This constraint prevents us {}from
obtaining the ac response, $\chi(\omega,F)$.

We now carry out a renormalization group analysis of Eq.(\ref{genfunc}) along
the lines
of Ref.\onlinecite{N1}. We rescale space and time as $x=bx^\prime$ and
$t=b^z t^\prime$, and integrate out modes of all frequencies in the
momentum shell between the (rescaled) momenta $b\Lambda$ and $\Lambda$,
where $\Lambda$ is the upper cutoff in momentum. The rescaling of the
fields $\Phi$ and $\hat\Phi$ are fixed in mean field theory by the requirement
that all
quadratic terms in the exponential have to be invariant under this
transformation, except the $\hat\Phi r\Phi$ term; the variation of this last
term is interpreted as a change in $F$ under renormalization.
Fixing the other quadratic terms in the exponential yields
$\hat\Phi=b^{-z-d/2}\hat\Phi^\prime$ and $\Phi=b^{z-d/2}
\Phi^\prime$, with the dynamic exponent $z$ fixed at 2 by comparing the
scaling of the $\partial_t$ and $\nabla^2$ terms. The consistency of the mean
field scaling is verified by ascertaining that all higher order operators are
irrelevant; this is indeed the case for $d>4$, which is therefore the upper
critical dimension. This upper
critical dimension is the same as for $F>F_T$.

At $d=4$, the entire function $C[\langle\phi(t_1)\rangle+\Phi(t_1)-(t_1
\leftrightarrow t_2)]$ becomes marginal. For $d<4$, the critical behaviour is
changed {}from its mean field form. As discussed in Ref.\onlinecite{N1}, for
$d<4$
the function $C$ splits up into a constant part $C_s$ that controls the
static distortions in the phase, and a functional operator $C(\phi)$ that
controls the dynamics. The two of these decouple {}from each other, and
scale differently.
For the dynamics above threshold, where the steady state solution is
periodic, it is necessary to preserve the invariance under the transformation
$\phi\rightarrow\phi+2\pi$. For $d<d_c$, the upper critical dimension,
fluctuations in $\phi$ scale in the same way as $\langle\phi\rangle=vt$;
this then requires $\Phi$ to be invariant under rescaling, as discussed in
Ref.\onlinecite{N1}. The operator whose coefficient is the constant part of
$C$,
$C_s$, is then {\it relevant\/}, leading to anomalous scaling of the
static correlations.

Below threshold, it is precisely these {\it static\/} correlations
that we are interested in. Since the mean phase $\langle\phi\rangle$
reaches a constant value for any fixed value of $F<F_T$ in steady state
(with $\langle\phi\rangle$ increasing with $F$), the periodicity of the
phase variable need not be preserved under the rescaling of the
renormalization group. Accordingly, we choose a  scaling
for the fields $\hat\Phi$ and $\Phi$ {\it different\/} {}from that of
Ref.\onlinecite{N1}:
\begin{equation}
\hat\Phi=b^{-d/2-z}\hat\Phi^\prime,\qquad\Phi=b^{2-d/2}\Phi^\prime.
\label{newscale}
\end{equation}
The scaling of the $\hat\Phi$ field is fixed by  requiring the invariance of
the $C_s$ term, which is unrenormalized by loop corrections. (This is
because all loop terms involve differences between the function $C$ at two
different values of its argument, which are not changed by an addition of
a constant $C_s$. Although it is conceivable that singular corrections
could affect this result, we assume here that this is not the case.)
As discussed in Ref.\onlinecite{N1}, any renormalization of the $\hat\Phi\Phi$
term {}from the operator
$C$ (as well as {}from higher order irrelevant operators) must be of the
form $\hat\Phi\partial_t\Phi$, and thus the $\hat\Phi\nabla^2\Phi$ term also
has
no loop corrections, fixing the scaling of $\Phi$. The scaling of the
$r\hat\Phi\Phi$ term yields a renormalized polarizability $\chi^\prime
=\chi/b^2$. Finally, the $(F-F_{MF})\hat\Phi$ term, which {\it does\/}
receive loop corrections, is relevant, so that the scaling of $F$ can also
be obtained by power counting: $(F-F_T)^\prime=b^{d/2}(F-F_T)$, thus
yielding the correlation length exponent
\begin{equation}
\nu=2/d\label{obtnu}
\end{equation}
and, {}from the scaling of $\chi$,
\begin{equation}
\gamma=4/d.\label{obtgamma}
\end{equation}
as mentioned in Eqs. (\ref{intronu}) and (\ref{introgamma}).

Although it is possible to obtain these exponents directly {}from a very
simplified model of CDWs\cite{Parisi}, the exponent $z$ {\it cannot\/} be
obtained from a similar analysis and, as we shall now see, is nontrivial.
A more detailed discussion of Ref.\onlinecite{Parisi} is given in Section V.
The exponent $z$ is not affected by the difference between the scaling here
and that of Ref.\onlinecite{N1}, although under
the scaling used here, the operator $C$ is dangerously irrelevant, flowing
towards a singular function of zero amplitude, with {\it constant\/} second
derivative $C^{\prime\prime}(\phi)$. (With the scaling of Ref.\onlinecite{N1},
$C$
flows towards a regular function of constant amplitude, with the same second
derivative.) The $O(\epsilon)$ result for $z$, obtained in
Ref.\onlinecite{N1}, depends on the second derivative of $C$, and is given by
\begin{equation}
z=z(F>F_T)=2-\epsilon/3+O(\epsilon^2).\label{obtz}
\end{equation}

The breakdown of two sided scaling for CDWs, with $\nu=2/d$ below threshold,
and $\nu=1/2$ above threshold, is now seen as the result of the different
static and dynamic behaviour of the system, rather than
being due to any fundamental difference between the properties above and
below threshold. Even above threshold, the static correlation length still
exists, and was indirectly obtained in Ref.\onlinecite{N1} as the length that
controls the finite size scaling behaviour. The static correlation
function above threshold varies as a simple power
law with distance, and  does {\it not\/} show any crossover at the static
(or dynamic) correlation length, so that the static correlation length
cannot be obtained directly.

\section{Numerical Results}

We now compare these results with those {}from numerical simulations.
Previous simulations have often been carried out with the continuous variable
dynamics of Eq.(\ref{FLR}), with $x$ discretized to a lattice. However, it is
possible to obtain much more accurate results with automaton models, which
may be viewed as a singular limit of Eq.(\ref{FLR}), in which the pinning
potential
at any site has very narrow and steep wells, so that $\phi(x;t)=\beta(x)
+2\pi m(x;t)$, with $m$ an integer variable. A further approximation is made by
discretizing time;\cite{ML} the dynamics of $m(x;t)$ is then
\begin{eqnarray}
m(x;t)&=&m(x;t-1)+1\qquad{\rm if}\,\,F+\nabla^2[2\pi
m(x;t-1)+\beta(x)]+h(x)>0\qquad\qquad\nonumber\\
&=&m(x;t-1)\qquad\qquad {\rm if} F+\nabla^2[2\pi m(x;t-1)+\beta(x)]+h(x)\leq
0.\label{discrtzn}
\end{eqnarray}
Here the $\nabla^2$ operator is the discrete Laplacian.  This automaton model
(and variants of it) are numerically
much more efficient than direct simulations of Eq.(\ref{FLR}). Since for small
$d$, the critical behaviour is
independent of the form of the pinning potential used, we shall rely mostly
on the automaton model simulations in this section, referring to continuous
dynamics only for comparison.

It is more convenient to express the dynamics in terms of a local ``curvature''
variable
\begin{equation}
c(x)=\sum_{y\in\langle xy\rangle} [m(y)-m(x)]
+{\rm Int}[\{F+\nabla^2\beta(x)+h(x)\}/2\pi]\label{curvature}
\end{equation}
where the sum over $y$ is restricted to sites neighbouring $x$.
The first term in this equation is the discretized form of $\nabla^2 m$. By
construction, $c(x)$ is an integer valued variable. {}From Eq.(\ref{discrtzn}),
we see that the dynamics can be expressed in terms of $c(x)$ as
\begin{eqnarray}
c(x)&\rightarrow& c(x)-n\nonumber\\
c(y)&\rightarrow& c(y)+1\,\,\forall y\in\langle xy\rangle\qquad{\rm if}\,\,
c(x)>0.\label{sandpile}
\end{eqnarray}
Here $n$ is the number of neighbours that the site $x$ has. We can view this
as a sandpile model\cite{BTW} if we treat $c(x)$ as the height of sand at
the site $x$; when this exceeds a preset threshold (zero in
Eq.(\ref{sandpile}))
one grain of sand falls off {}from $x$ to each of its neighbouring sites. As
$F$ is increased, ${\rm Int}[(F+\nabla^2\beta(x)+h(x))/2\pi]$ increases by 1
whenever its argument crosses an integer. {}From Eq.(\ref{curvature}), we see
that increasing $F$ towards threshold is equivalent to dropping grains of sand
randomly on different sites of the system. There is, however, a slight
difference in the manner in which sand is added as compared to standard
sandpile models: since the randomness {}from $\nabla^2\beta(x)+h(x)$ is
{\it quenched\/}, sand grains are dropped in a random order on the different
sites of the lattice one by one, with {\it no\/} site being revisited twice
in a cycle. When every site has been visited exactly once, the whole cycle is
repeated in the same order. In contrast to this, in standard sandpile models
the addition of each sand grain is an independent random process, and there are
no
correlations between the sites chosen. For a sufficiently large system, where
the repeating cycle is very long, it is reasonable to expect the difference
between cyclic and uncorrelated addition of sand grains to be
inconsequential.\cite{foot6}

Another difference between CDWs and usual sandpiles is that  open
boundary conditions are normally used for sandpiles, whereas periodic boundary
conditions
are used for CDWs. With periodic boundary conditions, a grain of sand that
exits the system at one boundary reenters at the opposite boundary.
Again, when the system size is much larger than the characteristic
avalanche size, this difference should not affect the
behaviour. \cite{foot4,foot5} In this section, we shall use numerical results
{}from CDW and sandpile models, {\it both\/} with periodic boundary
conditions, to compare with our analytical results. Comparisons with earlier
simulations\cite{alanref,Tang} on sandpiles with {\it open\/} boundary
conditions will be discussed in subsequent sections.

Figure 1 shows a finite size scaling plot for the polarization $P(F,L)$
(measured {}from the initial state $m=0$ at $F=0$) for $d=2$ as a function of
the
driving force $F$ and the linear size of the system, $L$. The finite size
scaling of the polarization should be of the form
\begin{equation}
P=|f|^{-\gamma+1}\hat P(L|f|^\nu)\label{fssP}
\end{equation}
where $f$ is the reduced force, $F/F_T-1$.
The collapse to a scaling form in the numerical data shown in Figure 1 is
not very good. This is probably due to the fact that the polarization in
any state involves the polarizability integrated {}from the initial to the
present state. Corrections to scaling might thus be expected to persist in the
polarization until the system is very close to threshold. Figure 2 shows a
similar scaling
plot for the polarizability, obtained by a numerical derivative of the
polarization. The data here scale better when $L|f|^\nu$ is large, but show
a large scatter very close to threshold due to numerical uncertainties.
Figure 3 shows a scaling plot for the polarization in the sandpile version of
the dynamics of Eq.(\ref{discrtzn}); the scaling form works much better here.
{}From the data in Figure 3, we obtain estimates for the critical exponents:
 $\gamma(d=2)=1.98\pm 0.03$ and
$\nu(d=2)=0.98\pm 0.03$, in agreement with Eqs.(\ref{obtgamma}) and
(\ref{obtnu}).

Simulations with
the continuous dynamics of Eq.(\ref{FLR}) have larger
numerical uncertainties. As detailed in Ref.\onlinecite{M1a}, a log-log plot of
the
polarization as a function of $f$ appears to yield an exponent $\gamma
=1.8\pm 0.15$, although the slight upward curvature seen as $F\rightarrow
F_T$ allows for the possibility that the asymptotic value of $\gamma$ is
indeed 2, as given by Eq.(\ref{obtgamma}). The collapse of the data to a single
scaling
function is also not very satisfactory, but can be used to obtain the
correlation length exponent $\nu$ as $1.0\pm 0.1$. A more accurate value
for $\nu$ is obtained {}from the width of the distribution of threshold
fields $F_T$ for various systems of size $L$ with different realizations of
randomness, since one expects finite size effects to become  important when the
characteristic size of an avalanche is of the order of the size of the system.
Fitting $\Delta F_T$ to the form $L^{-\nu}$  yields $\nu=1.01\pm
0.03$.\cite{M1a} The results are thus consistent
with Eq.(\ref{obtnu}).

For three dimensional systems, the numerical uncertainties in the
behaviour of the polarizability are larger. Here we show only the results
for simulations on the sandpile model (with periodic boundary conditions)
for which, as in two dimensions, the scaling is better. Figure 4 shows a
scaling plot of the polarizability, {}from which we obtain $\gamma=1.31\pm
0.08$
and $\nu=0.68\pm 0.04$. Since it is possible to
determine the location of the threshold field very accurately, one can
also plot the polarization {\it at\/} threshold as a function of
system size. As shown in Figure 5, $P_s(F_T,L)$ appears to scale as
$L^{0.57}$, consistent with Eqs. (\ref{obtnu}) and (\ref{obtgamma})
for $d=3$. (The error bars shown in Figure 4 are {}from the statistical
scatter in the data; owing to the small range in $L$, this is too conservative
an estimate for the true uncertainties.) The exponent $\nu$ can be obtained
independently {}from the scatter in the threshold field to be
$\nu=0.68\pm 0.02$, as shown in Figure 6.

{}From the duration of an avalanche as a function of its size, we obtain the
dynamic exponent $z$. The size $s$ of an avalanche, defined as the total
number of sites participating in it, scales as $l^d$, where
$l$ is its linear extent, while $z$ is defined through $t(l)\sim l^z$, so
that $t(s)\sim s^{z/d}$. Figures 7(a) and 7(b) are log-log plots of $t(s)$
versus
$s$ for CDWs in 2 and 3 dimensions respectively. {}From these we obtain
$z(d=2)=1.32\pm 0.04$ and $z(d=3)=1.65\pm 0.06$. This agrees very
well with our result $z(F<F_T)=z(F>F_T)$; $z(F>F_T)$ can be found {}from
the velocity exponent $\beta$ through $z(F>F_T)=2\beta$, and $\beta$ is
known independently {}from numerical simulations above threshold to
be $\beta(d=2)=0.64\pm 0.03$ and
$\beta(d=3)=0.81\pm 0.03$.\cite{M1b,Myers,Automaton}

\section{Distribution of Avalanches}

We now turn to the distribution of avalanches of different sizes in the
critical region. As discussed earlier, there are differences in the forms
for the continuous variable and automaton models. This is because the
total increase in the phase $\phi$ at any site is bounded above by $2\pi$
in the continuous variable models. No such constraint exists for automata,
for which several avalanches of roughly the same size are grouped together
into a single large avalanche. This may lead to different number distributions
for avalanches in the continuous and automaton models. We first deal with
the avalanche distribution for automaton models, returning later in this
section to continuous variable models.

We conjecture a scaling form for the number distribution of
avalanches\cite{M1a}:
\begin{equation}
n(s;f)ds={1\over{s^{\kappa/d}}}\hat n(s|f|^{\nu d}){{ds}\over
s}\label{avldistr}
\end{equation}
where $n(s;f)dsdf$ is the total number of avalanches of size between $s$
and $s+ds$ that occur in a  unit volume of the system when the reduced force
$(F/F_T-1)$ is changed {}from $f$ to $f+df$. This equation defines the
exponent $\kappa$. (Here we have assumed that the same $\nu$ is associated with
the characteristic size of the avalanches as well as the finite size effects.)

We define the moment of an avalanche as the total change in polarization that
results {}from it. This is related to the change in the phase at every site in
the avalanche by $\delta P=\sum_{i=1}^s (\delta\phi_i)$. For
continuous dynamics, where $\delta\phi_i\leq 2\pi$ for all the sites, the
moment scales as the size of the avalanche. For automaton
models, the two are related by a nontrivial exponent:
\begin{equation}
\delta P\sim s^\Gamma.\label{moment}
\end{equation}
Equations (\ref{avldistr}) and (\ref{moment}) can be used to obtain an
exponent identity relating $\kappa$ and $\Gamma$ to $\gamma$. The total
change in polarization upon changing the force by $df$ is found by
integrating Eq.(\ref{avldistr}) to be of the form
$\Delta P\sim f^{(\kappa-\Gamma d)\nu} df$.
Comparing with Eq.(\ref{suscep}), we obtain\cite{M1a}
\begin{equation}
\gamma=(\Gamma d-\kappa)\nu.
\end{equation}
Substituting Eqs.(\ref{obtgamma}) and (\ref{obtnu}) in this equation yields
\begin{equation}
\Gamma d-\kappa=2.\label{Gamma}
\end{equation}

Figure 8(a) shows a finite size scaling plot of the distribution of avalanches
for the automaton model in two dimensions. By fitting to the form of
Eq.(\ref{avldistr}), we
obtain
\begin{equation}
\kappa(d=2)=0.36\pm 0.03.\label{k2}
\end{equation} Figure 8(b) shows a similar plot for $d=3$,
{}from which we can obtain
\begin{equation}
\kappa(d=3)=1.00\pm 0.06.\label{k3}
\end{equation}
The values of $\nu$ that are obtained {}from
these scaling plots are $\nu(d=2)=0.98\pm 0.03$ and $\nu(d=3)=0.68\pm 0.03
$, consistent with
Eq.(\ref{obtnu}) (as well as the scatter in $F_T$).
The scaling of the moment of an avalanche
with its size can be used to obtain numerically
\begin{equation}
\Gamma(d=2)=1.15\pm 0.05\label{th2}
\end{equation}
and
\begin{equation}
\Gamma(d=3)=1.00\pm 0.02\label{th3}
\end{equation}
which, together with the numerical values for $\kappa$, are consistent with
Eq.(\ref{Gamma}).
In the limit as $F\rightarrow F_T$, a power law distribution for the
avalanches is obtained. If $\hat n(x)$ is not a singular function of
its argument $x$ for small $x$, the limiting form of the distribution
can be used to obtain $\kappa$ and $\Gamma$. If, however, $\hat n(x)$ scales
as $x^{(\alpha-1)/2d\nu}$ for small $x$, the apparent power law in the
avalanche distribution would change {}from $\kappa$ to
$\kappa-(\alpha-1)/2\nu$, while leaving $\Gamma$ unchanged. A singular form for
$\hat n(0)$ would imply
that the {\it total\/} number of avalanches generated by a small increase
in $F$ (which is dominated by small avalanches) would be singular as $F_T$ is
approached; we have verified numerically that this is not the case for the
automaton models.

It is interesting
to note that Eq.(\ref{Gamma}) has been derived earlier by Majumdar and
Dhar\cite{Majumdar} for sandpile models, by a
completely different method that relies only on properties of clusters
{\it at\/}  the critical point.\cite{foot7} Since sandpiles at the critical
point are
current driven, $J=0^+$, instead of being tuned by a driving force $F$, any
prefactor in $n(s)$ that is singular in $f$ would not be seen in the scaling.
The fact that Eq.(\ref{Gamma}) and the result of Majumdar and Dhar are
identical can thus be taken as a proof that $\hat n(0)$ is not singular. Our
results in Eqs. (\ref{k2}) through (\ref{th3}) are also consistent with
numerical simulations on sandpile
models in two and three dimensions.\cite{alanref}

As mentioned in the previous section, the dynamic exponent $z$ can be
found numerically {}from the scaling of the duration of an avalanche with
its size; the results in both two and three dimensions are fairly close
to a first order truncation of the $\epsilon$-expansion result of
Eq.(\ref{obtz}). Even at the critical point, the duration of the
avalanches can be used to obtain $z$; this allows us to compare with the
result for two-dimensional sandpiles\cite{Majumdar} that $z=5/4$, which
agrees with our numerical results. For three dimensions, there is
no exact calculation of $z$ at present to compare our numerical results
with. However, previous simulations on (a transformed version of) the
sandpile model with open boundary conditions\cite{alanref1} have yielded
$z(d=3)=1.62\pm  0.01$, in agreement with our results.

For continuous variable models, we have seen that the bound on the change
in phase at any site in an avalanche constrains $\Gamma_c$ to being 1.
(The subscript $c$ here denotes continuous variable dynamics.)
{}From Eq.(\ref{Gamma}), we then obtain $\kappa_c=d-2$, so that both the
exponents involved in the avalanche distribution are determined.\cite{foot8}
This
result is only true if the driving force is increased {\it infinitesimally\/}.
In practice, the force is increased in small steps; no matter
how small the step size, one presumably eventually crosses over into a regime
where successive avalanches at the same site trigger within a single force
step,
and the exponents change to those of the automaton models. Although
the analytical treatment of Section III is applicable only for continuous
variable dynamics, a proper analysis of the singular part of the
{\it non-linear\/} response to finite increases in $F$  might yield the
automaton values for the exponent $\Gamma$; we have so far been unable to
carry this out. Note that the numerical results in Eqs.(\ref{k2})
through (\ref{th3}) are  not very far {}from the trivial values $\Gamma=1$ and
$\kappa=d-2$; in fact, in three dimensions they are consistent to within
the numerical uncertainties.

Instead of triggering avalanches by increasing the driving force, it is
possible to do  so by applying small random `kicks' to the {\it phases\/} at
different sites, which may be considered to mimic the effect of thermal
noise. Remarkably, for small thermal noise, the characteristic linear
extent of an avalanche still scales with the distance {}from the zero
temperature threshold force, but with a {\it different\/} correlation
length exponent:
\begin{equation}
\xi_{th}\sim(F_T-F)^{-\nu_{th}}
\end{equation}
with $\nu_{th}\neq\nu$. The `thermal' correlation length exponent has been
found numerically by Myers\cite{Myers} to be $\nu_{th}\approx 0.50$ in two
dimensions. As shown in Figure 9, the same result is true for $d=3$ as
well. A proper understanding of this phenomenon is still lacking; however,
it is possible that considerations about the equilibrium behaviour (at
finite temperature) in the {\it absence\/} of a driving force may be applicable
here. At low temperatures, one might expect the presence of a driving
force to produce an overall forward drift of the CDW, while still preserving
the equilibrium properties locally. In equilibrium, the dynamics can
be expressed in terms of correlations in the pinning potential. This is
in contrast to our zero temperature analysis, where the physically
relevant correlation function $C$ is related to force-force
correlations.\cite{interf2} The potential-potential correlation function
is related to the function $C$ integrated twice; as shown in
Ref.\onlinecite{Ledou}, in equilibrium it is necessary for this to be periodic
in its argument.
Changing $C$ by a constant $C_s$ is therefore no longer allowed in
equilibrium. Setting $C_s$ to zero implies that the static correlations
no longer scale differently {}from the dynamic correlations, and
$\nu_{th}=1/2$.
Further work on this, especially on crossovers {}from equilibrium to
nonequilibrium properties, is clearly required.

\section{Discussion}

Most of the work on sandpile models has been on the properties at the
critical point. However, Tang and Bak\cite{Tang} have considered sandpile
models away {}from the critical point; through a combination of scaling
arguments and numerical simulations, they have obtained results similar to
some of ours.  For instance, by considering the diffusive
nature by which sand grains propagate through the system, they obtain
the exponent identity $\gamma/\nu=2$, which is the same as {}from
Eqs.(\ref{obtgamma}) and (\ref{obtnu}). With the assumption that the
distribution of avalanches is independent of system size for sufficiently
large systems, and the implicit assumption that the moment of an avalanche
is proportional to its size, they obtain an exponent identity connecting
$\gamma$ and $\kappa$, which is a special case of the general result derived
in the previous section. The first of these two assumptions is equivalent to
$\hat n(0)$ not being singular; we have seen in the previous section that
this is correct, both by comparing our scaling law to that of Majumdar and
Dhar\cite{Majumdar} and by direct numerical verification. The second
assumption, that $\Gamma=1$, is {\it not\/} correct, even though it is
approximately valid for $d=2$ which is the case Tang and Bak focus on (and
actually much better for $d=3$).  A few other results that
they obtain are for the behaviour of transients, and cannot be compared with
our work. The authors have also noted that if the critical behaviour in
these nonequilibrium models were to be like that of equilibrium
statistical mechanical systems, various  exponent identities relating
$\beta$ to other quantities would result.
These exponent identities can be compared with our results, and are found
{\it not\/} to be valid.

Surprisingly, the value obtained by them for
$\nu$ through numerical simulations  is $\nu(d=2)=0.7$, which disagrees
completely with our analytical and numerical result $\nu(d=2)=1$. It is
not clear what this difference is due to, although a similar value of
$\nu$ ($\nu=0.76$) has apparently been observed in other simulations as well in
which the sandpile is perturbed in a manner different {}from
ours\cite{Carlson}.  As discussed in the previous section, with different types
of kicks on the sandpile (that may or may not have physically meaningful
analogs in the continuous variable models) $\nu$ can be very different; this
might be the source of the discrepancy. Our simulations on sandpile models,
detailed in the previous section, yield $\nu(d=2)\approx 1$.

A related problem to CDWs, that of the depinning of interfaces in random
media, has
been analyzed in the moving phase by methods similar to those used
for CDWs.\cite{interf,interf2} Unlike the case for CDWs, the statics and
dynamics
do not behave differently; physically, this is because as the interface
moves forward it experiences different regions of impurities and is thus
unable to build up anomalous static correlations. Formally, this is seen
by the absence of an unstable direction at the fixed point of $C$ when the
appropriate boundary conditions are taken. Since this separate scaling  of
static and dynamic correlations is the source of the difference
between $\nu(F>F_T)$ and $\nu(F<F_T)$ for CDWs, two-sided scaling will be
valid for interfaces. The relation $\gamma/\nu=2$  can be verified to be still
true by the procedure in this paper.

Parisi and Pietronero\cite{Parisi} have considered a model for CDWs which is
equivalent to choosing the pinning potentials $V(\phi)$ at different sites to
be of the form
$V(\phi;x)=h(x)\phi-2\pi h(x)\sum_n\theta(\phi-\beta(x)-2\pi n)$, so that the
phase at any site experiences a {\it constant\/} retarding force {}from the
pinning potential, equal to $-h(x)$. In addition to this, there is the elastic
force from the neighbouring sites, and the external driving force. (Although
they have a $\theta$-function constraint that prevents the total force at any
site from being negative, we have seen that when threshold is approached
monotonically
the phases at all the sites always move forward, so that this constraint is
inconsequential for the critical behaviour.) Although this model
correctly yields $\gamma=4/d$ and $\nu=2/d$, the dynamic exponent $z$ is
wrong, being trivially equal to 2. The reason for this partially correct
result can now be understood: after averaging over the randomness, their
model corresponds to the one we have considered here, with the dynamic part
of the correlation function, $C(\phi)$, set to zero, but with $C_s\neq 0$.
This {\it does\/} constitute a fixed point, at which loop corrections {}from
$C(\phi)$
are absent, so that $z=2$. However, this fixed point is unstable to the one
that we have analyzed in this paper, so that the model that they have
considered constitutes a pathological limit for pinning potentials. In
particular, it is possible to replace the vertical drop in $V(\phi)$ at
$\phi=\beta+2\pi n$ with a very steep section of non-zero width.
The critical behaviour of this model has been found numerically\cite{M1b} to
the same as for the other models that we have considered here, although the
width of the critical region above threshold can be seen through a perturbation
expansion to vanish as the width of the steep sections grows smaller.\cite{N2}

As mentioned in Section III, due to the restriction that the driving
force must be monotonically increasing in time, it is not possible to
obtain the ac response $\chi(\omega,F)$ {}from our analysis. It is not clear at
present if some variation of our approach may prove adequate for this
purpose. It is important to remember that the universality that exists for
various different models with regard to the monotonic approach to threshold
is considerably limited for $\chi(\omega,F)$ even in low dimensions.

In this paper, we have analyzed the critical behaviour of charge density waves
as the depinning threshold is approached monotonically {}from below. Our
success in this was due to the fact that, for this special path to threshold,
the existence of many metastable states was argued not to invalidate the
perturbation expansion we carry out. Although for any particular problem one
must consider
afresh whether such a perturbation expansion is valid, it may be possible to
use similar techniques for other systems.

\centerline{\bf Acknowledgements}

We would like to thank Daniel Fisher, Satya Majumdar and Chao Tang for useful
discussions.
O.N. is supported by a Junior Fellowship {}from the Society of Fellows at
Harvard University, and wishes to acknowledge the hospitality of the Institute
for Theoretical Physics in Santa Barbara where part of this work has been
completed.

\vfill\eject
\begin{figure}
\caption{
Attempted finite-size scaling of the polarization $P$ for the CDW
automaton in $d=2$ with best fits $\gamma \approx 2$ and
$\nu \approx 1$. No set of exponents gives a single curve,
due to the large finite size effects and the arbitrary
constant in the definition of the polarization. Symbols
indicate the size of the system, while the numbers in
parentheses indicate the number of samples averaged over.
}
\label{fig1}
\end{figure}

\begin{figure}
\caption{
Scaling of the polarizability $\chi^{\uparrow}$ (numerical
derivative of the data in Fig. 1) in the $d=2$ CDW
automaton model with fitted exponents $\gamma=2\pm 0.15$
and $\nu=1\pm 0.1$. The scatter at low scaled fields
$L{|f|}^\nu$ is due to statistical fluctuations.
}
\label{fig2}
\end{figure}

\begin{figure}
\caption{
Scaling of the polarization in the $d=2$ sandpile automaton
model.  The scaling collapse is much better than in the
$d=2$ CDW model. {}From the range of exponents for which the
scaling collapse is within finite-size errors and statistical
uncertainties, we find $\gamma=1.98\pm0.03$ and $\nu=0.98\pm0.03$,
in agreement with the analytical results in the text.
}
\label{fig3}
\end{figure}

\begin{figure}
\caption{
Scaling of the polarizability in the $d=3$ sandpile automaton
model.
{}From the range of exponents for which the
scaling collapse is within finite-size errors and statistical
uncertainties, we find $\gamma=1.31\pm0.08$ and $\nu=0.68\pm0.04$,
in agreement with the analytical results of the text.
}
\label{fig4}
\end{figure}

\begin{figure}
\caption{
Polarization at threshold for the $d=3$ sandpile
automaton with periodic boundary conditions as a function
of linear size $L$.
Error bars represent $1\sigma$ statistical uncertainties.
The fitted slope in the figure indicates
only statistical errors and are an underestimate of the
true uncertainty in the ratio $(\gamma-1)/\nu$.
{}From such data, we estimate $(\gamma-1)/\nu = 0.57 \pm 0.08$.
}
\label{fig5}
\end{figure}

\begin{figure}
\caption{
Relative sample-to-sample fluctuations $\Delta F_T/F_T$ in
the threshold field $F_T$ as a function of linear size for
the $d=3$ sandpile model with periodic boundary conditions.
{}From the fitted slope, the finite-size correlation length
exponent $0.68\pm 0.02$ is obtained.
}
\label{fig6}
\end{figure}

\begin{figure}
\caption{
Numerical derivatives of the logarithm of the duration $t$ of
an avalanche with respect to the logarithm of the volume $s$,
for (a) the $d=2$ CDW model and (b) the $d=3$ sandpile model.
The approach to a constant at large $s$ gives the scaling
$t \sim s^{z/d}$, with $z=1.32\pm 0.04$ in $d=2$ and
$z=1.65\pm 0.06$ in $d=3$.
}
\label{fig7}
\end{figure}

\begin{figure}
\caption{
Scaled avalanche distributions seen at different fields for
(a) 100 samples of size $512^2$ in the $d=2$ CDW model and
(b) 110 samples of size $128^3$ in the $d=3$ sandpile model.
Avalanches result {}from the (adiabatically) increasing field.
We find $\kappa = 0.36 \pm 0.03$ and $\nu = 0.98\pm 0.03$ for
$d=2$ and $\kappa = 1.00 \pm 0.06 $ and $\nu = 0.68\pm 0.03$ for $d=3$.
}
\label{fig8}
\end{figure}

\begin{figure}
\caption{
Scaled distribution of avalanches for the $d=3$ sandpile
automaton model at fixed field (total height). Avalanches are
in response to individual ``thermal kicks'' which increase
the phase at one location (by redistributing the curvature
variable).  The scaling collapse is within errors for
$\kappa=0.99\pm 0.03$ and $\nu=0.50\pm 0.02$.
}
\label{fig9}
\end{figure}


\begin{references}
\bibitem[1]{Fisher} D.S. Fisher, Phys. Rev. Lett. {\bf 50}, 1486 (1983)
and Phys. Rev. B{\bf 31}, 1396 (1985).

\bibitem[2]{foot} This is only correct if one neglects various
effects like thermal fluctuations and dislocations; in real materials,
these either round out the sharp phase transition or drive it first order.
[See, {\it e.g.\/}, S.N. Coppersmith, Phys. Rev. Lett. {\bf 65}, 1044 (1990),
Phys. Rev. B{\bf 44}, 2887 (1991); S.N. Coppersmith and A.J. Millis, Phys.
Rev. B{\bf 44}, 7799 (1991).]

\bibitem[3] {Oldnum} P.B. Littlewood, Phys. Rev. B{\bf 33}, 6694 (1986);
H. Matsukawa, J. Phys. Soc. Jpn. {\bf 57}, 3463 (1988); S.N. Coppersmith
and D.S. Fisher, Phys. Rev. A{\bf 38}, 6338 (1988).

\bibitem[4]{Sibani} P. Sibani and P.B. Littlewood, Phys. Rev. Lett. {\bf 64},
1305 (1990).

\bibitem[5]{M1a} A.A. Middleton and D.S. Fisher, Phys. Rev. Lett. {\bf 66},
92 (1991) and Phys. Rev. B{\bf 47}, 3530 (1993).

\bibitem[6]{M1b} A.A. Middleton, Ph.D. thesis, Princeton University.

\bibitem[7]{Myers} C.R. Myers, Ph.D. thesis, Cornell University; C.R. Myers and
J.P. Sethna, (unpublished).

\bibitem[8]{Automaton}A.A. Middleton, unpublished; see also A.A. Middleton,
O. Biham, P.B. Littlewood and P. Sibani, Phys. Rev. Lett. {\bf 68}, 1586
(1992).

\bibitem[9]{N1} O. Narayan and D.S. Fisher, Phys. Rev. Lett. {\bf 68}, 3615
(1992) and
Phys. Rev. B {\bf 46}, 11520 (1992).

\bibitem[10]{M2} A.A. Middleton, Phys. Rev. Lett. {\bf 68}, 670 (1992).

\bibitem[11]{Parisi}G. Parisi and L. Pietronero, Europhys. Lett. {\bf 16}, 321
(1991).

\bibitem[12]{BTW} P. Bak, C. Tang and K. Wiesenfeld, Phys. Rev. Lett. {\bf 59},
381 (1987).

\bibitem[13]{alanref} S.S. Manna, Physica A{\bf 179}, 249 (1991).

\bibitem[14]{Tang} C. Tang and P. Bak, Phys. Rev. Lett. {\bf 60}, 2347 (1988).

\bibitem[15]{Majumdar} S. Majumdar and D. Dhar, Physica A{\bf 185}, 129 (1992).

\bibitem[16]{F2} We thank Daniel Fisher for pointing this out to us.

\bibitem[17]{Fukuyama} H. Fukuyama and P.A. Lee, Phys. Rev. B{\bf 17}, 535
(1978); P.A. Lee and T.M. Rice, Phys. Rev. B{\bf 19}, 3970 (1979);
L. Sneddon, M.C. Cross and D.S. Fisher, Phys. Rev. Lett. {\bf 49}, 292
(1982).

\bibitem[18] {convention} The velocity exponent is often called $\zeta$ in
previous references.

\bibitem[19]{sompol} H. Sompolinsky and A. Zippelius, Phys. Rev. B{\bf 25},
6860 (1982);
A. Zippelius, Phys. Rev. B{\bf 29}, 2717 (1984).

\bibitem[20]{ML} The discretization of time in unit steps results in a constant
force being indistinguishable {}from an ac force whose frequency is an integer.
This gives rise to mode-locking in the sliding state, as discussed in
Ref.\onlinecite{Automaton}. However, the discreteness of time does not affect
the
statics.

\bibitem[21] {foot6} This is the case for generic initial conditions for
$c(x)$. For instance, if the initial configuration is chosen to be $m=0$,
the second term in Eq.(\ref{curvature}) results in $c(x)$ being random.
However, if we choose $c(x)$ to be identically zero initially, the
behaviour with cyclic addition of sand grains is atypical.

\bibitem[22] {foot4} For $F>F_T$, where an infinite avalanche runs across the
entire system, the different boundary conditions give rise to very
different behaviour. The phase transition to a moving state that is seen
with periodic boundary conditions is destroyed with open boundary
conditions, where the system  expels an appropriate number of sand grains
to bring it back to its critical point.

\bibitem[23] {foot5} Note that both the quenched randomness in the addition
of sand grains as well as the different boundary conditions imposed result
in different {\it finite size corrections\/} for CDWs than for sandpiles,
even below threshold.

\bibitem [24]{foot9} We assume here that the avalanche clusters are compact.

\bibitem[25] {foot7} The notation they use is slightly different {}from
ours: they use exponents $\tau_d=1+[\kappa-(\alpha-1)/2\nu]/d$ and
$\tau_s=1+(\tau_d-1)/\Gamma$. Using the result {}from Eq.(\ref{obtnu})
that $\nu=2/d$, their result agrees with Eq.(\ref{Gamma}) only if
$\alpha=1$. (Although their result is expressed for the specific
case of $d=2$, it can be generalized straightforwardly.)

\bibitem[26] {alanref1} A.J. Guttmann and R.J. Bursill, J. Stat. Phys.
{\bf 59}, 1 (1990). See S.N. Majumdar, Phys. Rev. Lett. {\bf 68}, 2329
(1992) for the mapping to sandpiles.

\bibitem [27]{foot8} Note that $\alpha_c$ is {\it not\/} equal to 1; thus the
number of avalanches at a fixed value of $f$ scales with sizes with the
exponent $[\kappa_c-(\alpha_c-1)/2\nu]/d=-(\alpha_c-1)/2d\nu$. It is this
exponent that is denoted as $\kappa$ in Ref.\onlinecite{M1a}.

\bibitem [28] {interf2} O. Narayan and D.S. Fisher (unpublished).

\bibitem [29]{Ledou} T. Giamarchi and P. LeDoussal (unpublished).


\bibitem [30]{Carlson} J. Carlson, E.R. Grannan, C. Singh and G.H. Swindle
(unpublished).

\bibitem [31]{interf} T. Nattermann, S. Stepanow, L.-H. Tang and H. Leschhorn,
J. Phys. II (France) {\bf 2}, 1483 (1992)

\bibitem [32]{N2} O. Narayan (unpublished).

\end{references}
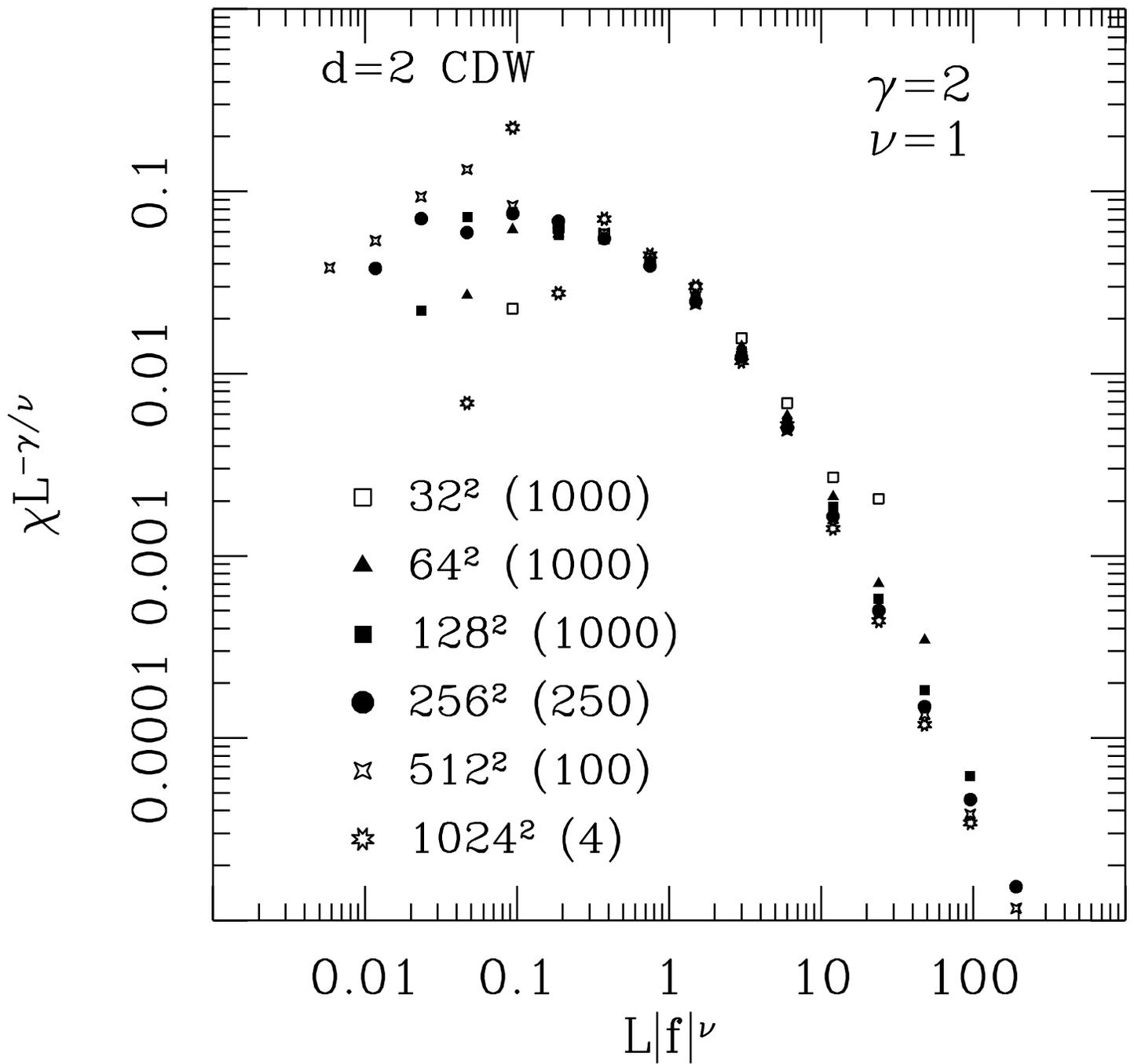
\end{document}